\documentclass[twocolumn,prl,showpacs,preprintnumbers,amsmath,amssymb,superscriptaddress,floatfix]{revtex4}
\usepackage{graphicx}

\begin{document}

\title{Lifetime Measurement of the 8$s$ Level in Francium}
\author{E. Gomez}
\affiliation{Department of Physics and Astronomy, SUNY Stony
Brook, Stony Brook NY 11794-3800 U.S.A.}
\author{L. A. Orozco}
\affiliation{Dept. of Physics, University of Maryland, College
Park, MD 20742-4111, U.S.A.}
\author{A. Perez Galvan}
\affiliation{Dept. of Physics, University of Maryland, College
Park, MD 20742-4111, U.S.A.}
\author{G. D. Sprouse$^1$}

\date{\today}

\begin{abstract}
We measure the lifetime of the $8s$ level on a magneto-optically
trapped sample of $^{210}$Fr atoms with time-correlated
single-photon counting. The $7P_{1/2}$ state serves as the
resonant intermediate level for two-photon excitation of the $8s$
level completed with a 1.3 $\mu$m laser. Analysis of the
fluorescence decay through the the $7P_{3/2}$ level gives $53.30
\pm 0.44$ ns for the $8s$ level lifetime.

\end{abstract}


\maketitle

We present in this letter a measurement of the $8s$ level lifetime
of francium; the heaviest and most relativistic of alkali atoms.
It is a test of the modern techniques of {\it ab initio}
calculations using many-body perturbation theory (MBPT)
\cite{johnson03,ginges04}. Fr is yet to be used in Parity
Non-conservation (PNC) measurements \cite{bouchiat97}, but work
towards that goal requires the understanding of the excited state
properties of the atom. It is important to ensure that our
quantitative understanding of the atomic structure of Fr is as
good as that of lighter alkali {\it e.g.} Cs where PNC experiments
have achieved resolution to extract weak force parameters
\cite{wood97}. Quantitative measurements on Fr and comparison with
theoretical calculations validate the same MBPT techniques used
for Cs and other atoms with a more relativistic atom where
correlations from the 87 electrons are large. The $8s$ state is
the preferred candidate for an optical PNC measurement; the dipole
forbidden excitation between the $7S_{1/2}$ ground state and the
first excited $8S_{1/2}$ state, becomes allowed through the weak
interaction.

The lifetime $\tau$ of an excited state is determined by its
individual decay rates, $1/\tau_i$, through the matrix element
associated with the $i$ partial decay rate. The connections
between lifetime, partial decay rates and matrix elements are:

\begin{equation}
\frac{1}{\tau_{i}}= \frac{4}{3}\frac{\omega^{3}}{c^{2}}\alpha
\frac{| \langle J \|r\| J' \rangle |^{2}}{2J'+1};~~~\frac{1}{\tau}
= \sum_i \frac{1}{\tau_i}, \label{lifetime}
\end{equation}
where $\omega$ is the transition energy divided by $\hbar$, $c$ is
the speed of light, $\alpha$ is the fine-structure constant, $J'$
and $J$ are respectively, the initial and final state angular
momenta, and $| \langle J \|r\| J' \rangle |$ is the reduced
matrix element \cite{grossman00b}. Equation \ref{lifetime} links
the lifetime of an excited state to the electronic wavefunctions
of the atom. The comparison of measurements with theoretical
predictions test the quality of the computed wavefunctions
specially at large distances from the nucleus due to the presence
of the radial operator.

The lifetimes of the low lying states in Fr are reaching a level
of precision comparable to that of the other alkalis
\cite{simsarian98,grossman00b,aubin04}. The atomic theory
calculations for these transitions
\cite{dzuba95,safronova99,dzuba01} predict the lifetimemes
measured with impressive agreement, strengthening the possibility
of a PNC experiment in a chain of francium isotopes.

We use the method of time correlated single photon counting to
obtain the lifetime of the $8s$ level in Fr in a magneto-optical
trap (MOT). We populate the $8s$ level with a two photon
transition, then we turn off the excitation suddenly and
observe the exponential decay through the fluorescence photons \cite{hoeling96}.

The production, cooling and trapping of Fr on-line with the
superconducting linear accelerator at Stony Brook has been
described previously \cite{aubin03a}. Briefly, a 100 MeV beam of
$^{18}$O ions from the accelerator impinges on a gold target to
make $^{210}$Fr (radioactive half life 3 min). We extract
$\sim$1$\times10^6$ francium ions/s out of the gold and transport
them 15 m to a cold yttrium neutralizer where we accumulate the Fr
atoms. We then close the trap with the neutralizer and heat it for
one second ($\sim$1000~K) to release the atoms into the dry-film
coated glass cell where they are cooled and trapped in a MOT. The
cycle of accumulating and trapping repeats every 20 s.

\begin{figure}
\leavevmode \centering
\includegraphics[width=3.1in]{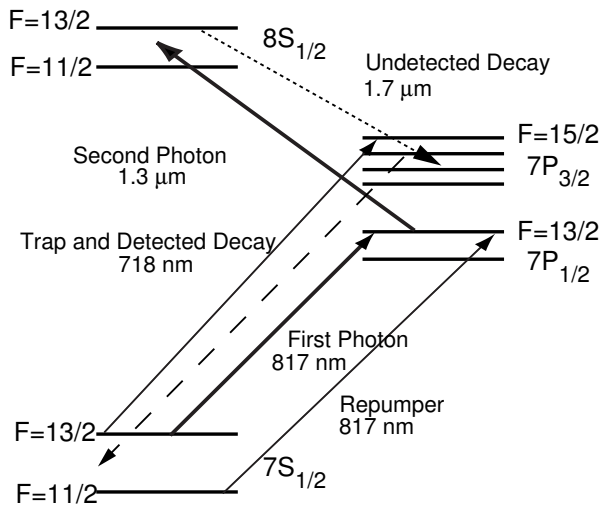}\caption{Energy levels of $^{210}$Fr. The figure
shows the trapping and repumping transitions (thin solid line),
the two photon excitation (thick solid lines), the fluorescence
detection used in the lifetime measurement (dashed line) and the
undetected fluorescence (dotted line). \label{levels}}
\end{figure}

Figure \ref{levels} shows the states of $^{210}$Fr relevant for
trapping and lifetime measurements.  A Coherent 899-21
titanium-sapphire (Ti:Sapph) laser operating at 718 nm excites the
trapping and cooling transition ($7S_{1/2},F=13/2 \rightarrow
7P_{3/2},F=15/2$). A Coherent 899-21 Ti:Sapph laser operating at
817 nm repumps any atoms that leak out of the cooling cycle via
the $7S_{1/2},F=11/2 \rightarrow 7P_{1/2},F=13/2$ transition. The
first photon for the $7S_{1/2} \rightarrow 8S_{1/2}$ transition
comes from a Coherent 899-01 Ti:Sapph at 817 nm, it populates the
$7P_{1/2}$, $F = 13/2$ state. The second photon at 1.3 $\mu$m
originates from an EOSI 2010 diode laser to excite the $7P_{1/2}
\rightarrow 8S_{1/2}$ transition.

A Burleigh WA-1500 wavemeter monitors the wavelength of all lasers
to about $\pm 0.001$ cm$^{-1}$. We lock the trap, first photon and
repumper lasers with a transfer lock \cite{zhao98}, while we lock
the second photon laser with the aid of a Michelson interferometer
that is itself locked to the frequency stabilized HeNe laser used
in the transfer lock.

The MOT consists of three pairs of retro-reflected beams, each
with 15 mW/cm$^2$ intensity, 3 cm diameter (1/e intensity) and red
detuned 31 MHz from the atomic resonance. A pair of coils
generates a magnetic field gradient of 9 G/cm. We work with traps
of $\approx 10^4$ atoms, a temperature lower than 300 $\mu$K, with
a diameter of 0.5 mm and a typical lifetime between 5 and 10 s.

\begin{figure}
\leavevmode \centering
\includegraphics[width=8.6cm]{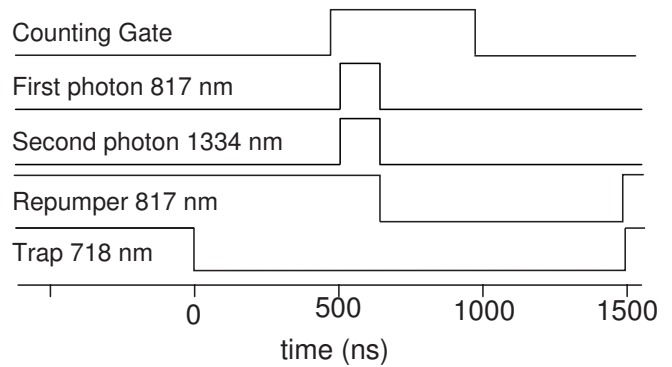}\caption{Timing diagram for the $8s$ level
excitation and decay cycle (100 kHz) \label{timing}}
\end{figure}

Figure \ref{timing} displays the timing sequence for the
excitation and decay cycle for the measurement. Both lasers of the
two photon excitation are on for 50 ns before they are switched
off, while the counting electronics are sensitive for 500 ns to
record the excitation and decay signal. The trap laser turns off
500 ns before the two-photon excitation. We repeat the cycle at
100 KHz.

We turn the trap light on and off with an electro optic modulator
(EOM) (Gs${\rm \ddot{a}}$nger LM0202) and an acousto optic
modulator (AOM) (Crystal Technology 3200-144). The combination of
the two gives an extinction ratio of better than 1600:1 after 500
ns. AOMs modulate the repumper and the first photon (817 nm) light
(Crystal Technology 3200), they have extinction ratios of 109:1
and 26:1  30 ns after the pulse turns off respectively. We couple
the 1.3 $\mu$m laser into a single mode optical fiber pass it
through a 10 Gbits/s lithium niobate electro-optic fiber modulator
(Lucent Technologies 2623N), then amplify it (Iphenix IPSAD1301)
and again modulate it with a second electro-optic fiber modulator
(Lucent Technologies 2623N); the result is an on-off ratio of
better than 1000:1 in a time of 20 ns.

A 1:1 imaging system (f/3.9) collects the fluorescence photons
onto a charge coupled device (CCD) camera (Roper Scientific,
MicroMax 1300YHS-DIF). We monitor the trap with the use of an
interference filter at 718 nm in front of the camera. A
beam-splitter in the imaging system sends 50$\%$ of the light onto
a photo multiplier tube (PMT) (Hamamatsu R636). An interference
filter at 718 nm in front of the PMT reduces the background light
other than fluorescence from the cascade through the $7P_{3/2}$
level decay back to the ground state $7S_{1/2}$.

After we turn off the excitation lasers, the atom returns back to
the ground level using two different decay channels (see
Fig.~\ref{levels}): First, by emitting a 1.3 $\mu$m photon it
decays back to the $7P_{1/2}$ state and fluoresces 817 nm light to
return to the $7s$ ground level. The second possible decay channel
is the $8s \rightarrow 7P_{3/2}$ transition followed by the decay
to the $7s$ ground level. The 1.7 $\mu$m fluorescence from the
first step of this decay is unobserved, but we detect 718 nm light
from the second part of the decay. With the known lifetime of the
$7P_{3/2}$ state, it is possible to extract the $8s$ level
lifetime from the cascade fluorescence decay.

We amplify (Ortec AN106/N) the current pulses from the photon
detections in the PMT.  We gate (EG\&G LG101/N) and send them to a
constant fraction discriminator (Ortec 934). The output starts a
gated time-to-amplitude converter (TAC) (Ortec 467) that we stop
with a fixed time delay pulse after the two-photon excitation.  We
use a multichannel analyzer (MCA) (EG\&G Trump-8k) to produce a
histogram of the events showing directly the exponential decay. A
pulse generator provides the primary timing sequence for the
measurement (Berkeley Nucleonics Corporation BNC 8010).

\begin{figure}
\leavevmode \centering
\includegraphics[width=3.1in]{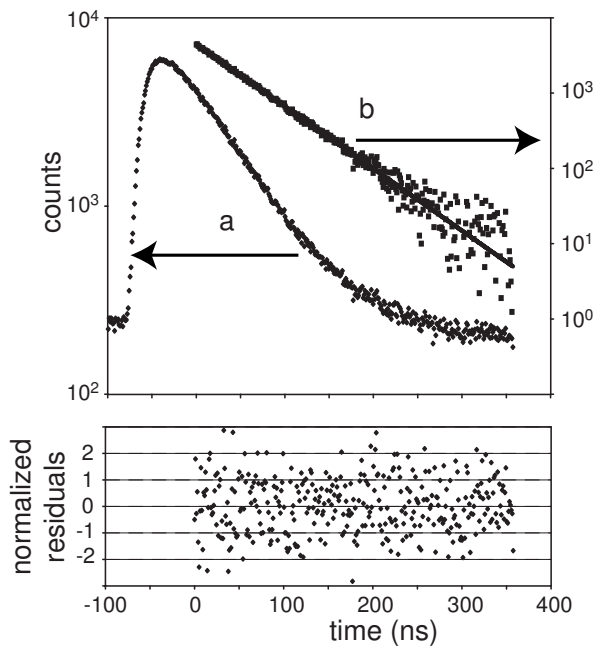}\caption{Cascading decay curve of the $8s$ level through the
$7P_{3/2}$ state in the MOT with fit and residuals. a: the arrival
time histogram data and b: data after the substraction of the
$7P_{3/2}$ decay and the background. The continuous line is the
fit. The lower plot shows the normalized residuals.\label{decay}}
\end{figure}

We take sets of data for about 1500 s, that are individually
processed, and fitted. The total number of counts in a set is
typically in the order of $3 \times 10^5$. Figure \ref{decay}
shows the accumulated decay of a set of data, together with the
exponential fit and the residuals.

We apply a pile-up correction, that accounts for the preferential
counting of early events \cite{oconnor84}. As low count rates keep
this correction small, we collect data with a small number of
fluorescence photons. We typically count one photon every 500
cycles. The correction alters the fitted lifetime by $+0.1\%$. We
perform a nonlinear least square fit and use an iterative
algorithm to find the fitting parameters that produce the smallest
$\chi^2$.

The decay signal $S_{8s}$ through the $7P_{3/2}$ state is a sum of
two exponentials \cite{aubin04} and a background $A_B$ with a
slope $A_S$, from the turn off of the trap. The fitting function
is:

\begin{equation}
S_{8s}= A_B +A_{S}t+A_{8s} \exp\left(-\frac{t}{\tau_{8s}}\right) +
A_{7p} \exp\left(-\frac{t}{\tau_{7p}}\right) \label{ffct},
\end{equation}
where $\tau_{7p}$ is the known lifetime of the $7P_{3/2}$ state
and $\tau_{8s}$, $A_B$, $A_{8s}$ and $A_{7p}$ are the fitting
constants. Figure \ref{decay} shows an example of a data set and
the fit. We start the fit 20 ns after both excitation lasers are
turned off. The fitting function describes the data well, and the
reduced $\chi_{\nu}^2$ of this particular decay is 1.11. A
discrete Fourier transform of the residuals shows no structure.
The average $\chi_{\nu}^2$ for all the data files used to obtain
the lifetime is $1.07 \pm 0.07$. A change (within our quoted
uncertainty)  on the calibration of the linearity of the MCA is
responsible for deviation from unity of the reduced chi squared.
The slope that we find is 0.02 counts in 500 channels for a
counting time of 1 second and it is the remanent of the trap
light. A fit to a file consisting of the sum of all files gives
consistent results both for the $8s$ lifetime and for the
$7P_{3/2}$ lifetime when this last one is left as a free
parameter.

We calculate the contribution to the uncertainty in the $8s$
lifetime from the $7P_{3/2}$ lifetime of 21.02(11) ns
\cite{simsarian98} using Bayesian statistics \cite{aubin04}. The
$7P_{3/2}$ state gives a Bayesian error of 0.15\%.

We do not observe any systematic effects depending on the start
and end points of the fit, the so called truncation error, beyond
the statistical uncertainty. We look for effects in the lifetime
from imperfect lasers turn off by leaving the first photon on
continuously. The change in the lifetime with the first photon off
or continuously on during the decay constraints the uncertainty
from imperfect lasers turn off to $\pm 0.07$ \%. The time
calibration of the pulse detection system contributes $\pm 0.01$
\% to the uncertainty. The TAC and MCA nonuniformity contribute
$\pm 0.11$ \% error in the $8s$ level lifetime and increases the
value of the $\chi_{\nu}^2$.

We study the effect of the initial state conditions on the
obtained lifetime by changing external parameters of the
measurement. We vary the power of the 817 nm first photon laser
and we observe no change in the measured lifetime. The time of
flight of the atoms can influence the measured 8s level as excited
atoms may leave the imaging region before they fluoresce. However,
the average velocity of the atoms in the MOT is less than 0.1 m/s
and the imaging region has a diameter of 1 mm. The time it takes
the atoms to traverse the imaging region is approximately $10^5$
times the measured $8s$ level lifetime.

The slope in the fitting function influences the value of the
obtained lifetime. Files with and without the second photon that
produces the background exponential decay give a consistent slope.
We compare the lifetime obtained by leaving the slope as a free
parameter or by fixing it to the background files value and obtain
an uncertainty contribution of $\pm 0.36$ \%.

The counting PMT is continuously on and detects light from both
the two photon excitation and the fluorescence light from the MOT.
We bound the possible saturation effects on the PMT by comparing
the average response of the PMT in photon counting mode with the
response of a fast photodiode not subject to saturation. W e find
a maximum contribution of $\pm 0.24\%$ to the overall uncertainty
from the PMT recovery.

We search for other possible systematic effects in the lifetime of
the equivalent level ($6s$) in Rb, given the complications of
working with Fr. These measurements are performed both in a vapor
cell and in a MOT . There can be collisional quenching or
radiation trapping in a gas of atoms that can modify the lifetime;
however, we find no evidence of change when we vary the number of
atoms from $10^3$ to $10^5$ in the Rb trap and we establish a
limit on radiation trapping from the Rb data of $\pm 0.01$ \%. We
have performed an extensive search for some additional magnetic
sensitivity: there is no change in the lifetime beyond the
statistical uncertainty when we change the gradient of the Fr MOT.
The detection of the cascaded photon reduces the possibility of
quantum beats \cite{hoeling96}. We establish a limit on magnetic
field effects of $\pm 0.11$ \% in the uncertainty of the Fr
measurement consistent with our work in Rb \cite{gomez04b}.

Table \ref{errorbudget} contains the error budget for the $8s$
level lifetime measurement. The statistical error dominates the
uncertainty of the measurement. We obtain a lifetime of 53.30
$\pm$ 0.44 ns for the $8s$ level of francium.

\begin{table}
\renewcommand{\arraystretch}{1.3}\centering
\begin{tabular}{lll}
&  Error [\%]\\ \hline

\hspace{0.5cm}Time calibration&$\pm$0.01 \\
\hspace{0.5cm}Bayesian error&$\pm$0.15\\
\hspace{0.5cm}TAC/MCA response nonuniformity&$\pm$0.11 \\
\hspace{0.5cm}Radiation trapping&$\pm$0.01 \\
\hspace{0.5cm}Imperfect laser turnoff&$\pm$0.07\\
\hspace{0.5cm}Magnetic Field&$\pm$0.11 \\
\hspace{0.5cm}Background slope&$\pm$0.36\\
\hspace{0.5cm}PMT response&$\pm$0.24\\

\hspace{0.5cm}Statistical error &$\pm$0.65\\ \hline
\hspace{0.5cm}{\bf Total} &{\bf$\pm$0.82}\\
\end{tabular}
\caption{Error budget for the $8s$ level lifetime measurement.}
\label{errorbudget}
\end{table}

\begin{figure}
\leavevmode \centering
\includegraphics[width=3.1in]
{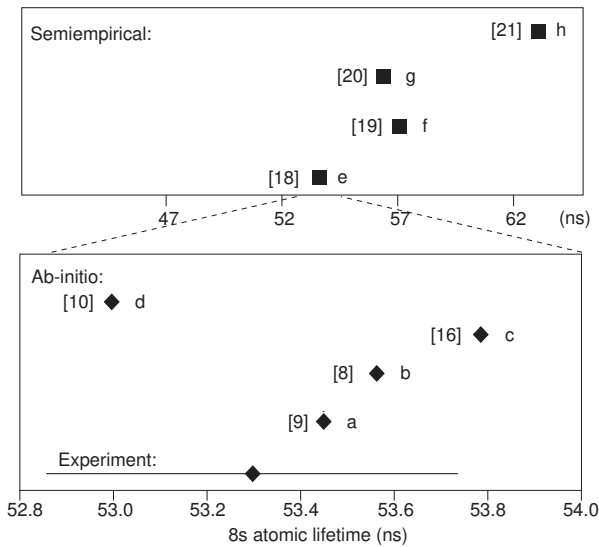} \caption{Comparison of the $8s$ level lifetime with
theory. The calculations are labelled with letters explained in
the text and with numbers that indicate the
reference.}\label{comparison}
\end{figure}

Figure \ref{comparison} compares the obtained $8s$ level lifetime
with theoretical calculations.  $a$ to $d$ are  \textit{ab initio}
MBPT calculations of the dipole matrix elements by $a:$ Safronova
\textit{et al.} \cite{safronova99}, $b:$ V. A. Dzuba \textit{et
al.} \cite{dzuba95},  $c:$ W. R. Johnson \textit{et al.}
\cite{johnson96}, and $d:$ V. A. Dzuba \textit{et al.}
\cite{dzuba01}. We calculate the lifetime with Eq.~\ref{lifetime}
from their predictions and measured transition energies
\cite{simsarian99}. $e$ to $h$ are semiempirical calculations:
$e:$ M. Marinescu \textit{et al.} \cite{marinescu98}, $f:$ C. E.
Theodosiou \cite{theodosiou94}, $g:$ E. Bi\'{e}mont \textit{et
al.} \cite{biemont98}, and $h:$ W. A. van Wijngaarden \textit{et
al.} \cite{wijngaarden99}. The scatter of results from the MBPT
calculations is small and they are all within one percent of our
result. The semiempirical methods are less accurate and they have
a broader scatter for their predictions (expanded scale in
Fig.~\ref{comparison}).

Our measurement establishes that the MBPT calculations of matrix
elements that contribute to the total lifetime of the state are
very good. They take into account the large relativistic effects
present in this heavy atom as well as the multiple correlations
from its 87 electrons. Their accuracy is vital for future
interpretation of PNC measurements. The agreement of theoretical
predictions over different species reinforces the interpretation
of PNC measurements in Cs which are now sensitive to the nuclear
weak force \cite{wood97}.%

Work supported by NSF. E. G. acknowledges support from CONACYT and
the authors thank the personel of the Nuclear Structure Laboratory
at Stony Brook for their support as well as J. Gripp, J. E.
Simsarian and B. Minford for equipment loans.


\vspace*{-0.2cm}
\begin{thebibliography}{10}

\bibitem{johnson03}
W.~R. Johnson, M.~S. Safronova, and U.~I. Safronova, Phys. Rev. A
{\bf 67},
  062106  (2003).

\bibitem{ginges04}
J.~S.~M. Ginges and V.~V. Flambaum, Phys. Rep. {\bf 397},  63
(2004).

\bibitem{bouchiat97}
M.-A. Bouchiat and C. Bouchiat, Rep. Prog. Phys. {\bf 60},  1351
(1997).

\bibitem{wood97}
C.~S. Wood, S.~C. Bennett, D. Cho, B.~P. Masterson, J.~L. Roberts,
C.~E.
  Tanner, and C.~E. Wieman, Science {\bf 275},  1759  (1997).

\bibitem{grossman00b}
J.~M. Grossman, R.~P.~Fliller III, L.~A. Orozco, M.~R. Pearson,
and G.~D. Sprouse,
  Phys. Rev. A {\bf 62},  062502  (2000).

\bibitem{simsarian98}
J.~E. Simsarian, L.~A. Orozco, G.~D. Sprouse, and W.~Z. Zhao,
Phys. Rev. A {\bf
  57},  2448  (1998).

\bibitem{aubin04}
S. Aubin, E. Gomez, L.~A. Orozco, and G.~D. Sprouse, Phys. Rev. A
{\bf 70},  042504
  (2004).

\bibitem{dzuba95}
V.~A. Dzuba, V.~V. Flambaum, and O.~P. Sushkov, Phys. Rev. A {\bf
51},  3454
  (1995).

\bibitem{safronova99}
M.~S. Safronova, W.~R. Johnson, and A. Derevianko, Phys. Rev. A
{\bf 60},  4476
   (1999).

\bibitem{dzuba01}
V.~A. Dzuba, V.~V. Flambaum, and J.~S.~M. Ginges, Phys. Rev. A
{\bf 63},
  062101  (2001).

\bibitem{hoeling96}
B. Hoeling, J.~R. Yeh, T. Takekoshi, and R.~J. Knize, Opt. Lett.
{\bf 21},  74
  (1996).

\bibitem{aubin03a}
S. Aubin, E. Gomez, L.~A. Orozco, and G.~D. Sprouse, Rev. Sci.
Instrum. {\bf
  74},  4342  (2003).

\bibitem{zhao98}
W.~Z. Zhao, J.~E. Simsarian, L.~A. Orozco, and G.~D. Sprouse, Rev.
Sci.
  Instrum. {\bf 69},  3737  (1998).

\bibitem{oconnor84}
D.~V. O'Connor and D. Phillips, {\em Time Correlated Single Photon
Counting}
  (Academic, London, 1984).

\bibitem{gomez04b}
E. Gomez, F. Baumer, A. Lange, L.~A. Orozco, and G.~D. Sprouse, to
be submitted
   (2004).

\bibitem{johnson96}
W.~R. Johnson, Z.~W. Liu, and J. Sapirstein, At. Data Nucl. Data
Tables {\bf
  64},  279  (1996).

\bibitem{simsarian99}
J.~E. Simsarian, W.~Z. Zhao, L.~A. Orozco, and G.~D. Sprouse,
Phys. Rev. A {\bf
  59},  195  (1999).

\bibitem{marinescu98}
M. Marinescu, D. Vrinceanu, and H.~R. Sadeghpour, Phys. Rev. A
{\bf 58},  R4259
   (1998).

\bibitem{theodosiou94}
C.~E. Theodosiou, Bull. Am. Phys. Soc. {\bf 39},  1210  (1994).

\bibitem{biemont98}
E. Bi{\'{e}}mont, P. Quinet, and V. van Renterghem, J. Phys. B
{\bf 31},  5301
  (1998).

\bibitem{wijngaarden99}
W.~A. van Wijngaarden and J. Xia, J. Quant. Spectrosc. Radiat.
Transf. {\bf
  61},  557  (1999).

\end{thebibliography}

%

\end{document}